\documentclass[pdflatex,sn-mathphys-num]{sn-jnl}

\usepackage{graphicx}%
\usepackage{multirow}%
\usepackage{amsmath,amssymb,amsfonts}%
\usepackage{amsthm}%
\usepackage{mathrsfs}%
\usepackage[title]{appendix}%
\usepackage{xcolor}%
\usepackage{textcomp}%
\usepackage{manyfoot}%
\usepackage{booktabs}%
\usepackage{algorithm}%
\usepackage{algorithmicx}%
\usepackage{algpseudocode}%
\usepackage{listings}%
\usepackage{rotating}

\theoremstyle{thmstyleone}%
\theoremstyle{thmstyletwo}%

\theoremstyle{thmstylethree}%

\begin{document}

\title[Article Title]{The Unbearable Weight: Scaling Models and Methods for UAV Audio Classification}

\author*[1]{\fnm{Andrew P.} \sur{Berg}}\email{berga2@g.cofc.edu}

\author[2]{\fnm{Qian} \sur{Zhang}}\email{zhangq@cofc.edu}

\author*[1]{\fnm{Mia Y.} \sur{Wang}}\email{wangy5@cofc.edu}

\affil*[1]{\orgdiv{Department of Computer Science}, \orgname{College of Charleston}, \orgaddress{\city{Charleston} \state{SC}, \country{USA}}}

\affil[2]{\orgdiv{Department of Engineering}, \orgname{College of Charleston}, \orgaddress{\city{Charleston} \state{SC}, \country{USA}}}

\abstract{As unmanned aerial vehicles (UAVs) become increasingly prevalent in consumer and defense settings, classifying them reliably from limited, modality-specific data is an urgent challenge. The dominant approach, large pretrained networks fully fine-tuned on task data, carries a substantial computational and memory weight that is hard to bear in resource-constrained UAV deployments, where edge inference and rapid retraining for emerging platforms are both required. This paper systematically scales across both model architectures and fine-tuning methods for UAV audio classification, asking when that weight is justified and when lighter alternatives prevail. Using a custom dataset of 3,100 audio clips spanning 31 drone classes, we evaluate transformer (ViT, AST) and convolutional (custom CNN, ResNet-18/152, MobileNet-V3-S/L, EfficientNet-B0/B7) backbones under full fine-tuning, classifier-only fine-tuning, and four parameter-efficient fine-tuning (PEFT) methods: SSF, IA3, OFT, and selective batch-norm tuning. All configurations are evaluated with 5-fold cross-validation across accuracy, training time, trainable-parameter share, and inference-time memory footprint. Selective batch-norm fine-tuning of EfficientNet-B7 with three-fold augmentations achieves the highest validation accuracy ($97.65\% \pm 0.30$) while updating under 0.5\% of model parameters. Across the sweep, lightweight CNNs consistently outperform transformers on both accuracy and efficiency. For UAV audio classification under data scarcity, scaling the method outperforms scaling the model.}

\keywords{UAV Classification, Audio Signal Processing, Parameter Efficient Fine Tuning, Transfer Learning, Model Scaling}



\maketitle

\section{Introduction}\label{sec1}
Unmanned Aerial Vehicles (UAVs), commonly called drones, pose security and intelligence threats to private and public interests. Defense and intelligence sectors increasingly need to do more than detect that an unknown UAV is present; they need to classify which platform it is, so that engagement, jamming, and forensic decisions can match the threat. Among the viable sensing modalities (radio frequency, radar, optical, audio), audio offers a particularly accessible trade-off: it is short-range but cost-effective and does not depend on lighting or active emissions from the target \cite{b47}. This paper focuses on UAV audio classification; future work will integrate visual and radar telemetry into the same pipeline.

UAV audio captures the acoustic signature of a drone in flight: the broadband noise of rotors and motors, the harmonics of propeller blade-pass frequencies, and aerodynamic effects induced by the airframe. It is not a recording of mission content or telemetry, but a recording of the platform itself. Different UAVs have characteristic spectral fingerprints determined by motor count, blade geometry, and frame design, which is why audio is useful for distinguishing one model from another rather than merely flagging that a UAV is present. Both indoor and outdoor recordings preserve these signatures; outdoor recordings additionally capture environmental factors such as wind, birdsong, and traffic noise.

A core difficulty for UAV audio classification is data scarcity. Capturing a representative dataset for every consumer or military UAV in active use is prohibitively expensive: physical platforms must be sourced, recorded across environments, and labeled, and the resulting dataset goes out of date as new drones reach market. Because labeled UAV audio is therefore inherently limited compared to general audio classification benchmarks, training large randomly-initialized deep networks from scratch is not viable; the model is overdetermined for the available data. The natural alternative is to leverage models pretrained on large, related corpora and adapt them to the UAV task. This adaptation, however, introduces its own cost: full fine-tuning of large pretrained backbones is computationally heavy, slow to retrain as new drones are added, and erases much of what the pretrained weights encoded. Parameter-efficient fine-tuning (PEFT) methods and targeted data augmentation are practical responses to this tension, but their interaction with model scale on small-data audio tasks is not well-characterized.

\begin{figure}[ht]
    \centering

    \includegraphics[width=1\textwidth]{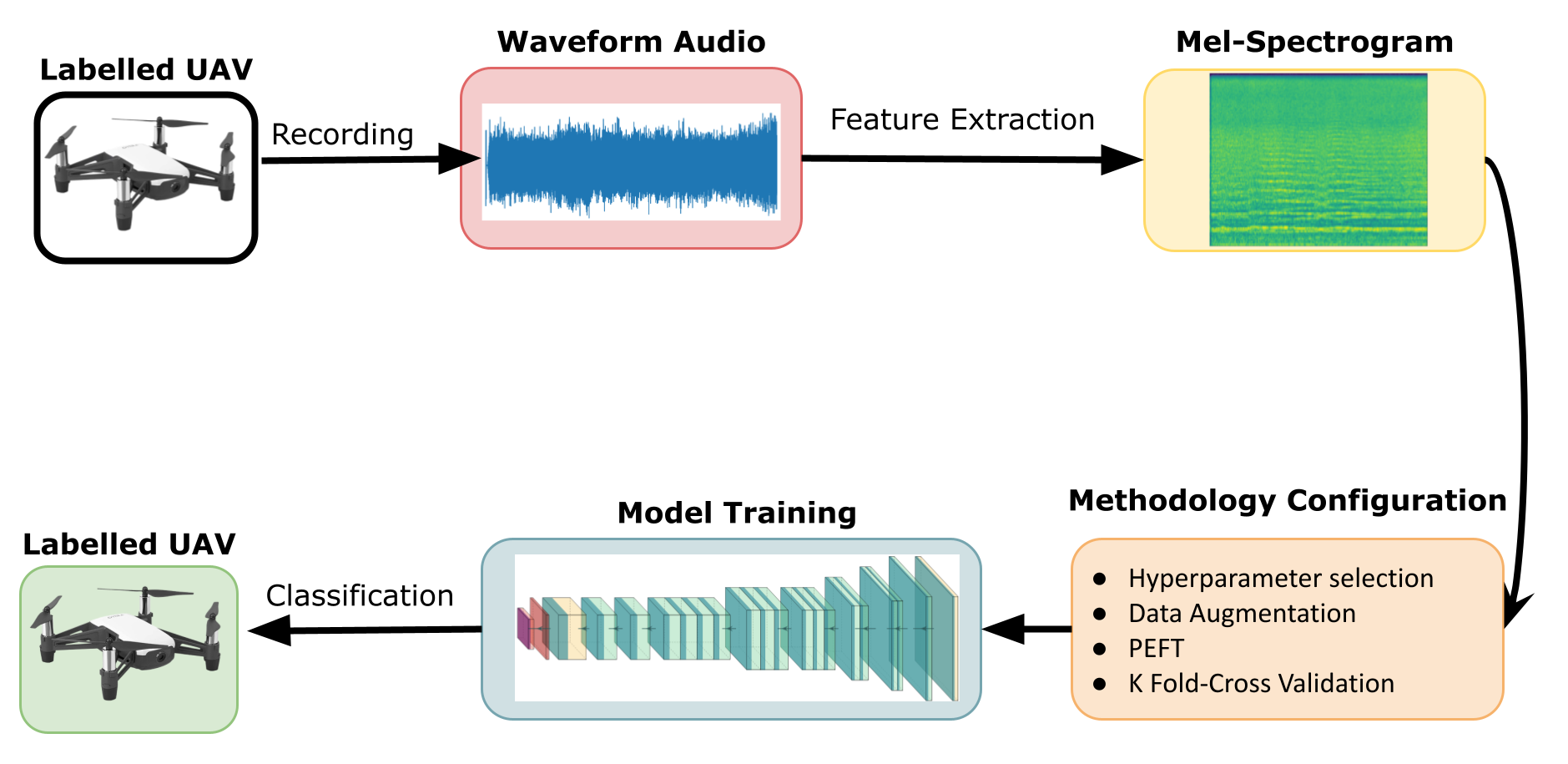}
    \caption{UAV Audio Classification System}
    \label{fig:trainig_system}
\end{figure}

This paper systematically scales across both model architectures and fine-tuning methods to characterize that interaction for UAV audio classification. We compare convolutional and transformer backbones at multiple sizes (custom CNN, ResNet-18/152, MobileNet-V3 small and large, EfficientNet-B0/B7, ViT-Base, AST) against full fine-tuning, classifier-only fine-tuning, and four PEFT methods (SSF, IA3, OFT, and selective batch-norm tuning), reporting accuracy, training time, trainable-parameter share, and inference-time memory footprint on a 3,100-clip, 31-class UAV audio dataset. The framing is the trade-off the title invokes: large pretrained models carry an ``unbearable weight'' of compute and memory in resource-constrained UAV settings, and the central question is when that weight is justified and when a smaller model or a lighter fine-tuning method is the better engineering choice. Figure \ref{fig:trainig_system} shows the high-level system from recording to classification.

\textbf{Three research questions follow directly from this framing:}
\begin{enumerate}
    \item Can fine-tuned transformers outperform fine-tuned CNNs on our 31-class UAV audio dataset, given the well-known data-scale advantage of transformers?
    \item What combination of model architecture, fine-tuning method, and data augmentation strikes the best accuracy-to-efficiency balance under data scarcity?
    \item Within each model family, how does scaling the backbone (e.g., EfficientNet-B0 vs. B7, ResNet-18 vs. -152, MobileNet small vs. large) shift the accuracy/efficiency trade-off when combined with PEFT?
\end{enumerate}

\section{Related Work}

\textbf{Previous Work:}
This paper is a continuation based on previous work presented in \cite{b1}. The prior paper explored an extremely small data approach of a 9-class dataset (4,500 seconds) and UAV audio classification, comparing transformers and convolutional neural networks (CNNs). The former concluded that the custom CNN approach had slightly better performance than the pre-trained audio spectrogram transformer (AST). The custom UAV audio dataset established in \cite{b22}\cite{Wang_dissertation} has since grown to the current 31-classes (15,500 seconds). The large leap in data scale presents an interesting look into real world data scaling for UAV classification.

\textbf{UAV Classification:}
Reference \cite{b47} explains the 4 major modalities for UAV classification: audio, optical, radar, and radio frequency (RF). For audio, the paper mentions, it is effective in short ranges, but is limited in noisy environments. Optical methods provide strong target classification, but depend on correct lighting and weather, and are still limited by range. Radar supports long-range capability, but is limited by radio interference. RF analyze UAV-emitted signals, but relies on the drone emitting signals and can face interferences. In reviewing these modalities, audio emerges as an accessible choice, both being practical and cost effective. This motivates our focus on deep UAV audio classification.

\textbf{Deep Learning \& Audio Classification:}
Deep Learning (DL), a subfield of machine learning (ML), is characterized by the use of deep neural networks to learn on a dataset. DL models are increasingly replacing ML approaches in audio classification, frequently beating state-of-the-art (SOTA) benchmarks. DL features a diverse range of architectures suitable for audio classification. The study \cite{b2} found the most common approach for audio classification to involve CNNs, using feature extraction techniques to transform the raw audio content using various methods like relative waves and mel-scale transformations. We narrow our model selection to CNNs and transformers. Another paper \cite{b3} compares pre-trained transformers and CNNs, indicating that transformers can significantly outperform the CNNs. The authors also indicate the strong usefulness of transfer learning.

\textbf{Transfer Learning:} Transfer learning (TL) takes a model typically pre-trained on a large dataset and is used as the initialization for a new, related task. By reusing the model weights from the previous task the newly fine-tuned model can leverage the previous knowledge of the pre-trained model. This reduces both training and dataset scarcity challenges constraints \cite{b25}. However, when domain-specific pre-trained networks are unavailable, cross-modal TL provides a practical alternative; wherein a model trained in one modality is later fine-tuned to a task separate from the original modality (e.g. image to audio classification) \cite{rethingingCNNs}. This is an effective way to simplify the training approach for many systems and achieves comparable performance to state of the art (SOTA) benchmarks, especially when a relevant model doesn’t exist in a given niche. This increases the range of viable pre-trained models.

\section{Methodology}
\subsection{Data Collection}
The drone data collection is still an on-going project \cite{b50} \cite{b51}. So far, for the past 5 years, drone audio data was collected from 31 distinct unmanned aerial vehicles (UAVs), as detailed in Table~\ref{tab:uav_data_31}. Each UAV contributed 100 five-second audio recordings, resulting in a total of 3,100 audio files and 15,500 seconds of raw flight audio. The dataset is perfectly balanced. These recordings encompass a broad spectrum of consumer, commercial, and custom-built UAV platforms and were conducted across both indoor and outdoor environments in three U.S. locations. The recordings capture the drone in flight (hover or short controlled flight near the microphone), so the acoustic content is the platform's rotor, motor, and aerodynamic signature rather than mission telemetry. Indoor recordings isolate this signature with minimal ambient noise; outdoor recordings layer it with environmental sounds (wind, birdsong, traffic) described in the Recording Sites subsection below.

\textbf{Drone Overview:} The dataset includes 28 quadcopters, one tricopter, and one hexacopter. Most of the quadcopters use a conventional X-frame design typical of consumer drone platforms. UAVs were sourced from manufacturers including DJI, Autel, Syma, Yuneec, UDI, Hasakee, Holystone, Hover, and two self-built drones. Notably, \textit{David Tricopter}, designed by David Windestal, is a custom-built tricopter with a 34-inch diameter, equipped with an AfroFlight Naze32 flight controller and weighing approximately 2.6 lbs. \textit{PhenoBee}, designed by Ziling Chen, is the largest UAV in the dataset, weighing approximately 23 kg with a 1.35-meter diameter, and operates using the Ardupilot framework with Cube Orange hardware.

\textbf{Recording Sites:} Data were collected in West Lafayette, Indiana; New Richmond, Indiana; and Charleston, South Carolina. Indoor recordings in Indiana were conducted in a university lab setting, while outdoor recordings were collected on a private farm in New Richmond. Charleston indoor data were collected in the Drone Lab at the Harbor Walk Campus of the College of Charleston, and outdoor recordings were captured from the rooftop of the aquarium parking garage. Environmental factors varied naturally during recordings and included wind, birdsong, traffic noise, and changing weather conditions.

\textbf{Recording Equipment:} From 2021 through 2023, audio data were captured using a MacBook Air with a 1.1GHz quad-core Intel Core i5 processor and 8GB of memory. Beginning in 2024, recordings used a MacBook Air with an Apple M3 chip and 16GB of memory. All data were recorded using the system's built-in microphones without external post-processing.

\subsection{Technical Environment}
\textbf{Software:}
The code is available at our public GitHub repository \cite{b4}
The model training runs are tracked using Weights \& Biases and are publicly available as well \cite{b5}. Ensuring full reproducibility of the experiments. All training runs were inside Docker containers using python version 3.11, CUDA 12.1.0, and Ubuntu 20.04.

This project utilized various machine learning frameworks; the following are listed with brief descriptions and references: PyTorch as the essential deep learning framework of choice \cite{b6} \cite{b7}; Transformers and PEFT packages\cite{b8}\cite{b9}; PyTorch lighting to simplify training loops \cite{b10}; Weights and Biases for experiment logging and hyperparameter searching \cite{b11}; Docker for server and version management \cite{b12}. Other python packages such as: Torchaudio \cite{b13}, Audiomentations\cite{b14}, TorchMetrics\cite{b15}, MatPlotLib \cite{b16}, SciKitlearn\cite{b17}, NumPy\cite{b18}, and Librosa\cite{b19} were used.

\textbf{Hardware:}
We ran all model experiments on a local server built with an AMD Threadripper 4950x, 128 Gbs of RAM, and two NVIDIA GeForce RTX 4090s; both equipped with 24 Gbs of VRAM. Notably, we chose to only train with one of the two GPUs, because the distributed runs took longer. In future work this may be a major bottleneck; however, for the current data scale it is not prohibitive. We aim to keep the model training as accessible as possible.

\subsection{Feature Extraction}
Feature extraction is common in audio classification \cite{b21} as it improves the performance of models in audio tasks, with strong precedent for use of mel-spectrograms, which were also used in previous UAV audio classification work \cite{b22}. Alternatives include mel-frequency cepstral coefficient (MFCC), mel-filter banks and (non-mel scale) spectrograms. The raw recorded UAV audio is recorded in waveform format (using a .wav file). This representation of a sound signal is a time-domain signal that represents how loud a sound (amplitude) is over time and is commonly used for playback to human ears. Spectrograms, however, a representation of how the frequency content changes over time, which is better suited for ML and DL models \cite{mu2021environmental}. A mel-spectrogram produces a representation of sound adjusted for human perception. While the mel-scale was initially proposed for pyschoacoustics, it is still a beneficial methodology for models to reduce dimensionality and emphasize frequency ranges.
TorchAudio \cite{b13} was used to compute the Mel-Spectrogram. The sampling rate was set to 16000, the FFT window size to 1024 (n-fft), overlap between frames to 128 (hop-length) and the number of mels to 128 (n-mels). The mel-spectrograms were computed during training time.

Feature extraction begins with a \textbf{Hanning window function}, defined as:

\[
w[n] = \tfrac{1}{2}\Big[1 - \cos\Big(\tfrac{2 \pi n}{N}\Big)\Big], \quad 0 \le n \le N-1
\]

where $w[n]$ is the window function and $N$ is the total number of samples in each window.
Here, zeroth indexing is used. The role of this window is to smooth the signal and reduce spectral leakage,
achieved through the $\big(1 - \cos(\tfrac{2\pi n}{N})\big)$ term.
This pre-processing step improves the accuracy of subsequent feature representations. After windowing, the \textbf{Short-Time Fourier Transform (STFT)} is applied.
For a continuous-time signal, it is defined as:

\[
\text{STFT}_x(t, \omega) = \int_{-\infty}^{\infty} x(\tau)\, w(\tau - t)\, e^{-j \omega \tau} \, d\tau
\]

Here, the raw signal $x(\tau)$ is localized in time by the window function $w(\tau - t)$,
and then projected onto the sinusoidal basis $e^{-j \omega \tau}$.
Computing the STFT across all frames produces a \textbf{spectrogram}, which shows how frequency content evolves over time. Next the \textbf{mel-scale} is often applied to the spectrogram, along with a log term.
The frequency mapping from Hertz ($f$) to mels ($m$) is given by:

\[
m(f) = 2595 \cdot \log_{10}\left(1 + \tfrac{f}{700}\right)
\]

\begin{figure}[ht]
    \centering

    \includegraphics[width=1\textwidth]{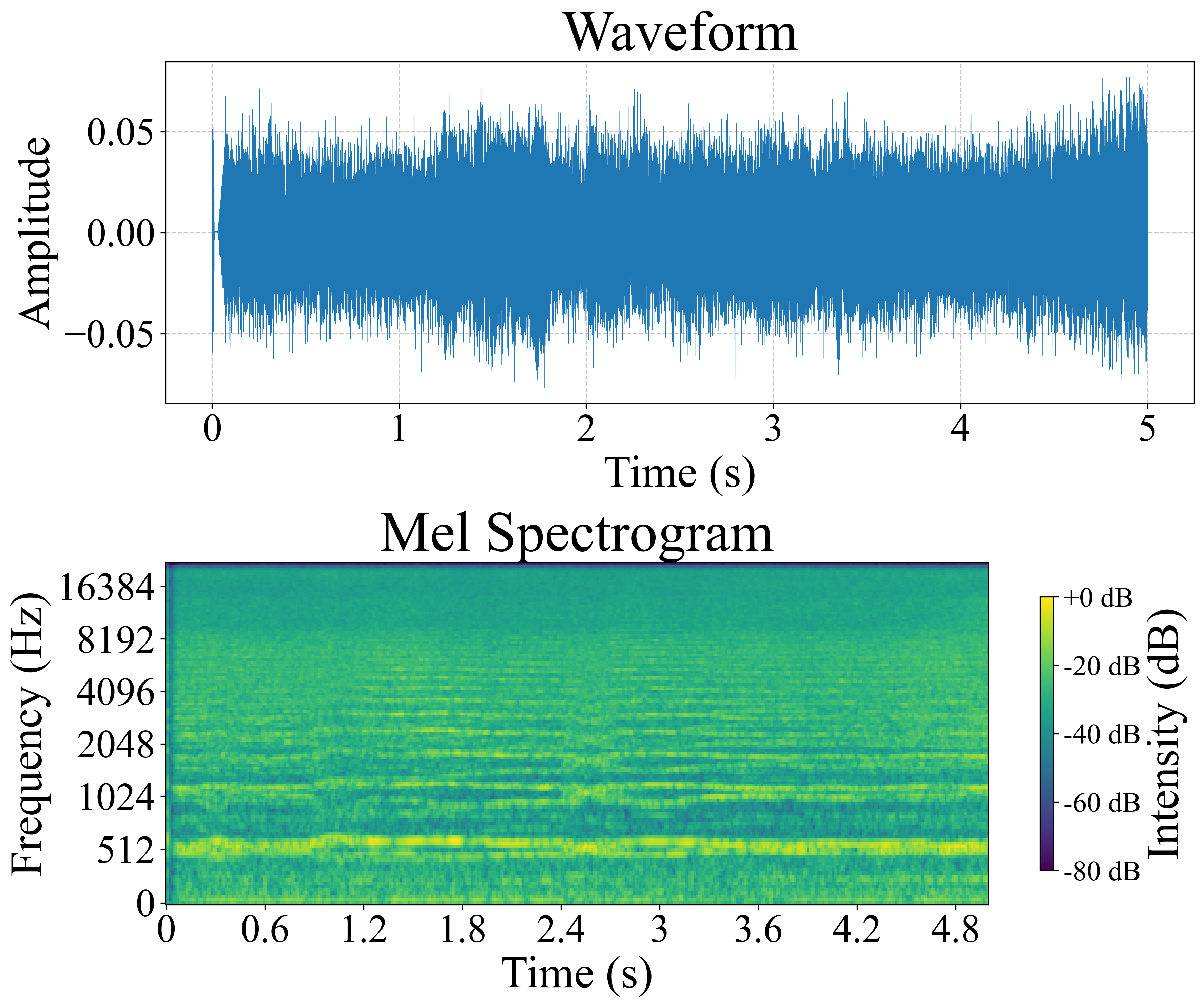}
    \caption{Audio Analysis of DJI Tello Drone; adapted from \cite{b1}}
    \label{fig:audio-analysis}
\end{figure}

This nonlinear transformation compresses higher frequencies while preserving finer detail in lower frequencies. In this case we use a log-scaled mel-spectrogram. A visual comparison of audio waveform and mel-spectrogram is given in Fig. \ref{fig:audio-analysis}.

\subsection{Data Augmentations}
\begin{figure}[ht]
    \centering

    \includegraphics[width=1\textwidth]{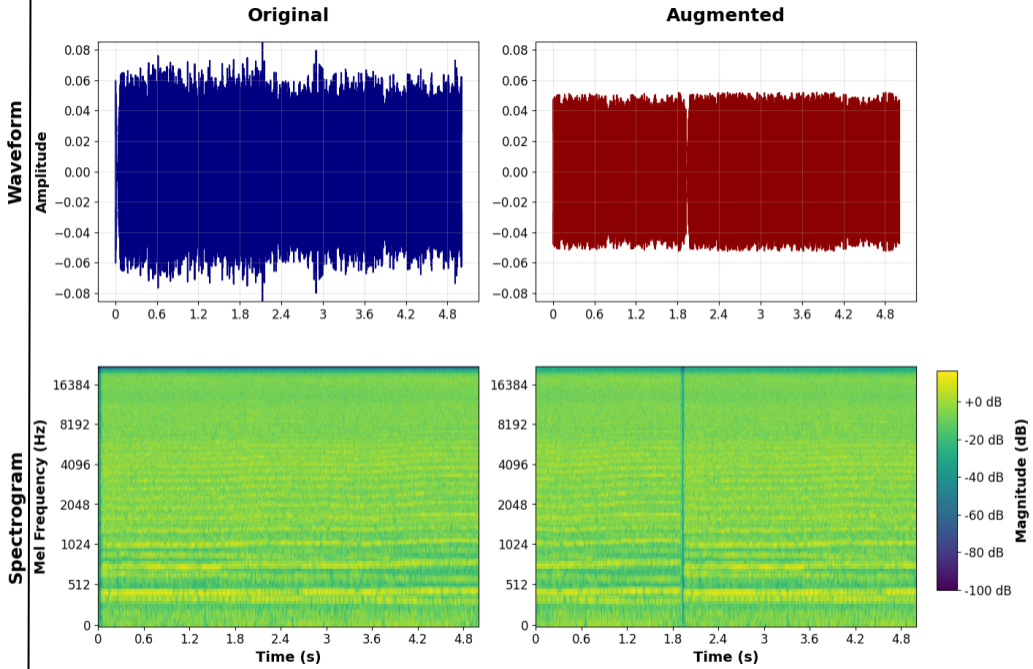}
    \caption{Comparison of Audio Augmentations}
    \label{fig:augmentation_comparisons}
\end{figure}

Data augmentations vary in their application and methods. Reference \cite{b23} narrows the scope of relevant UAV audio augmentations distortion, pitch and time, and noise. These mimic real-world audio variations that our custom dataset might have missed. The usage of data augmentations is especially important to address the inherent data scarcity of UAV audio classification. Time stretch and harmonic (sin) distortion are both applied during training implemented using the Audiomentations package \cite{b14}. For each original training sample, three deterministic augmentations were generated using a composition of time stretch and harmonic distortion. inflating the training set by 3 times while keeping the validation set unaltered, as the augmented data may not entirely reflect real world audio. The data augmentations were applied during training time. A comparison between augmented waveform and spectrogram is visualized in Fig. \ref{fig:augmentation_comparisons}. A full exploration of available waveform augmentations is available in public Google Colab notebook \cite{b24}.

\subsection{Training}

We utilize the pre-training/fine-tuning paradigm of model training. A pre-trained model, defined as any model that has been trained on a dataset or task before we train it. We leverage these pre-trained models using various fine-tuning strategies to effectively learn for a new smaller dataset and downstream task\cite{b25}. Essentially a model's previous learned representations are repurposed and applied to new data. The most common way to adapt a pre-trained model to a dataset is to train all of the model’s parameters. This is feasible for large enough datasets, but is slow and has massive forgetting from the pre-trained weights. We call this \textbf{full fine-tuning}. An alternative to this is linear probing, or as we refer to it in our experiments, \textbf{classifier fine-tuning}, which leaves only the final classifier block trainable \cite{b29}. This is an ablation study tactic used to judge a model's pre-trained generalization to a new task. This fine-tuning technique is not reliable as it does not have the effective size to generalize effectively, but is included nonetheless to  benchmark against other methods of fine-tuning.

\subsection{PEFT}
Parameter efficient fine tuning (PEFT) is a general term that describes any training method that leverages pre-trained neural networks, where only a small subset of parameters are updated \cite{b48}. We differentiate between the PEFT methods that can be used with transformers and those that can not. For large networks like large language models (LLMs) PEFT can significantly speed up fine-tuning.

\textbf{OFT (Orthogonal Fine Tuning):} Initially proposed for the fine-tuning of text to image generation models, OFT aims to capture what \cite{b30} refers to as “hyper-spherical energy” of a model. This boils down to the higher importance of a parameter’s orthogonal (diagonal) values. OFT does this by multiplying a model’s weights by a scaled and rotated orthogonal weight. We implemented this using the python PEFT package from Hugging Face \cite{b9}. In theory this could be implemented for both CNNs and Transformers, however, we opted to apply it only to its intended structure of transformers.

\textbf{Ia3 (Infused Adapter by Inhibiting and Amplifying Inner Activations):} Proposed to rival in-context learning in large language models (LLMs) \cite{b31}, Ia3 introduces new learnable vectors that target the query, key, value attention mechanism as well as after the non-linear activation function after the attention mechanism. It is implemented using the PEFT package. This PEFT method is only suitable for transformers, as it targets the attention mechanism.

\textbf{SSF (Scale Shift Fine Tuning):} SSF was proposed to rival the more computationally expensive PEFT methods \cite{b32}. This PEFT method is purely additive, meaning that it simply introduces a learnable scaling and additive vector to the targeted weights. We chose to implement this PEFT method from scratch and have included it in our repo, its code can be found directly in our repository at \cite{b33}. This PEFT method is model agnostic and we use it for both CNN and Transformer models.

\textbf{Batch Norm Fine-tuning:} Batch norm Fine-tuning is a selective PEFT
method that was initially proposed in \cite{b34}, which showcases solid
performance whilst only training on batch norm layers in CNNs. We implemented
this PEFT method ourselves, simply only training all batch norm parameters and
keeping the classifier layer as trainable. This PEFT method is only viable for
CNN models as transformers do not (typically) use batch norm layers.

\textbf{Applicability summary.} The PEFT methods divide cleanly by model family. OFT and IA3 target the transformer attention block and are applied only to ViT and AST. Batch-norm fine-tuning targets the per-channel normalization layers that appear throughout convolutional backbones and is applied to ResNet, MobileNet-V3, and EfficientNet. SSF is architecture-agnostic and is applied to all pretrained backbones: for transformers it scales and shifts the layer-normalized outputs of each transformer block, and for CNNs it scales and shifts the post-batch-norm convolutional feature maps. Our custom CNN is randomly initialized and is therefore reported only under full fine-tuning. Classifier fine-tuning is included for the remaining backbones as an ablation, and full fine-tuning is reported as the upper bound on adaptation.

\subsection{Models}
A motivating question to this study is: are Transformers or CNNs better for the
use case? The main delimiter, often, is the scale of the data. Dually, the
inductive bias that CNNs have to image and audio modalities makes them effective
in small data tasks. However, when trained on a high enough quantity of data the
transformer can surpass CNN performance on key benchmarks. This is noted in the
vision transformer white paper \cite{b35} which explores a purely attention
based image classification network. We can partially overcome the issue of lack
of data by using pre-trained models/networks that already have “learned
representations” of related data. Despite possibly being pre-trained on
different modalities; the pre-training/fine-tuning paradigm is empirically
proven to be more effective than random weights \cite{b36}.

\begin{figure}[ht]
    \centering

    \includegraphics[width=1\textwidth]{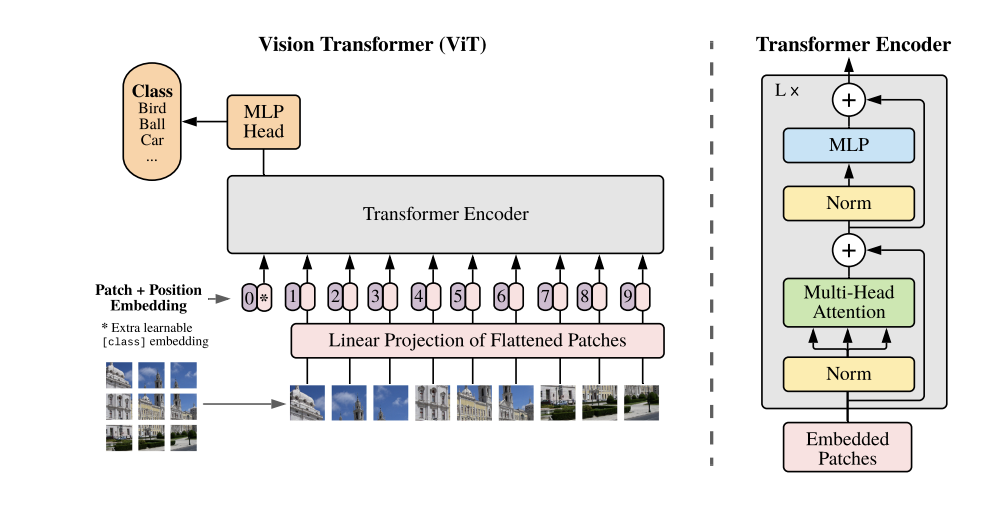}
      \caption{ViT visualization adapted from \cite{b35}}
    \label{fig:vit}
\end{figure}

\textbf{ViT (Vision Transformer):} Published in 2021 \cite{b35} and pre-trained
on JFT-300M \cite{sun2017}, which is a private and custom Google dataset with
over 300 million labeled images. The vision transformer is pre-trained on
JFT-300M and fine-tuned to smaller datasets like ImageNet. This is a key
paradigm in the current atmosphere in transformers and key training paradigm in
this paper. The vision transformer innovates on the original
transformer from \cite{vaswani2023}. Introducing a friendlier encoding
mechanism, leveraging fixed-sized patches of the input images and linearly embed
them into the encoding blocks, as well as introducing an extra learnable
classification token to the embedding space; which is critical to the problem of
image classification, and by proxy audio classification. The ViT architecture is
shown in Fig. \ref{fig:vit}

The ViT paper notes that ``Vision Transformer has much less image-specific
inductive bias than CNNs,'' and that this deficit is overcome only through
pretraining on JFT-300M, a corpus of roughly 300 million labeled images.
Critically, the authors report that ViT only begins to outperform comparable
CNNs once pretraining reaches the order of 14 million images; below that
threshold, CNNs retain the advantage. Our UAV audio dataset is roughly four
orders of magnitude smaller than this crossover point, and the ViT checkpoint we
use is pretrained on image data rather than audio. We therefore enter the
experiment expecting CNNs to retain their small-data advantage on UAV audio; our
experimental results, reported below, confirm this expectation.

We implemented a pre-trained version of ViT using the
``google/vit-base-patch16-224'' checkpoint from Hugging Face. This specific
version was fine-tuned using ImageNet.

\textbf{AST (Audio Spectrogram Transformer):} Audio Spectrogram Transformer (AST) is a pre-trained transformer, which is based on the ViT model, integrating and merging the pre-trained ViT b-16 weights into its own model. Further it is trained on the ImageNet and Audioset datasets. It is a stark example of both pure attention based audio classification and cross modality transfer learning, using both image and audio data to train on.

The Audio Spectrogram Transformer (AST) \cite{b37} is an audio classification transformer pre-trained on the ImageNet \cite{b38} and AudioSet \cite{b39} Dataset, leveraging cross modal transfer learning. AST adapts its model architecture from the popular vision transformer (ViT) \cite{b35}.

\begin{figure}[ht]
    \centering

    \includegraphics[width=1\textwidth]{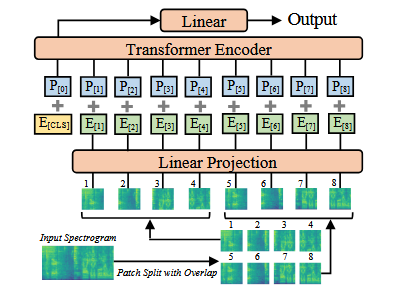}
      \caption{AST visualization adapted from \cite{b37}}
    \label{fig:ast-model}
\end{figure}

As shown in Fig. \ref{fig:ast-model} the AST uses patchified spectrograms as inputs to the encoder. Notably, there is no decoder structure as seen in natural language processing (NLP) transformers, but instead a fully connected linear layer for classification. We implemented a pre-trained version of AST using the MIT/ast-finetuned-audioset-10-10-0.4593 checkpoint from Hugging Face. This version was only trained using Audioset.

\begin{figure}[ht]
    \centering

    \includegraphics[width=1\textwidth]{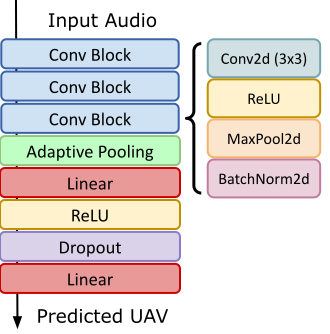}
    \caption{Custom CNN Architecture}
    \label{fig:custom-cnn}
\end{figure}

\textbf{Custom CNN:} Previously used and tested in \cite{b1}, our standard CNN, pictured in Fig. \ref{fig:custom-cnn}. Uses 3 stacked convolutional layers with a final fully connected sequence to act as the classifier. This model is not pretrained on any data. Our custom CNN acts as a point of comparison against the other pre-trained models in this paper.

The model's architecture is as follows: mel-spectrogram input shape into 3 convolutional blocks consisting of conv2d operation with a kernel size of 3 and padding of 1, a ReLU activation function, max pooling(2D) and a batch normalization(2D) layer. After that a simple dense classification layer that consists of adaptive pooling, a fully connected layer dynamically calculated based on the feature size, a dropout layer, and finally a classifying linear layer set to an output shape of 31.

\textbf{ResNet:} Introduced in \cite{b40} ResNet introduced the residual block into the deep learning field, completely revolutionizing how deep models can be stacked. Simply, the residual block, or skip connection, re-adds a model’s input to a learned weight, this innovation avoids massive forgetting on large/deep neural networks. When it was published, ResNet achieved State of the art (SOTA) performance on the ImageNet benchmark. We implement ResNet v1.5 available at \cite{b41} experimenting on two model sizes: resnet18 and resnet152. We have slightly tweaked the model's layers to agree with our input and output shapes, replacing the first convolutional layer to accept a grayscale input shape.

\textbf{MobileNet} Originally introduced in 2017, MobileNet \cite{b42}, didn’t aim for the top performance on ImageNet, but rather for efficiency and for the smallest model possible. This aligns with UAV classification applications, which can be applied on edge/mobile devices. We implement the pre-trained updated MobileNet v3 \cite{b43} using both the large and small model variants, sourced from TorchVision \cite{b44}.

\textbf{EfficientNet:} Introduced in 2019 \cite{b45} EfficientNet aimed to strike a balance between a model's scaled depth, width, and depth. In finding an optimal balance between the scaling of CNNs, EfficientNet sits between the extremes of the efficiency of MobileNet and the larger models. It is implemented using the base 0 and base 7 sizes \cite{b46}.

\section{Experiments \& Results}

All models were evaluated with 5-fold cross validation, with a 80\% training and 20\% test split for each fold. All models were trained under the same computational budget to ensure comparability. Hyper parameters are chosen as defaults from prior work \cite{b1}. Data augmentations were confined to the training set. For all model training runs the batch size was 8, with 2 gradient accumulation steps; giving an effective batch size of 16. Models were trained for 20 epochs each. Gradient clipping was enabled, with gradients rescaled whenever their L2 norm exceeded 1.0. Model checkpointing was enabled, monitoring the validation loss.

For CNNs used a learning rate of 1e-3 with a step learning rate scheduler applied every 10 epochs reducing by a gamma of 0.1. Transformer models used a learning rate of 1e-4 with a cosine annealing scheduler with a minimum learning rate of 0 and maximum number of iterations set to 50.

\begin{figure}[ht]
    \centering

    \includegraphics[width=1\textwidth]{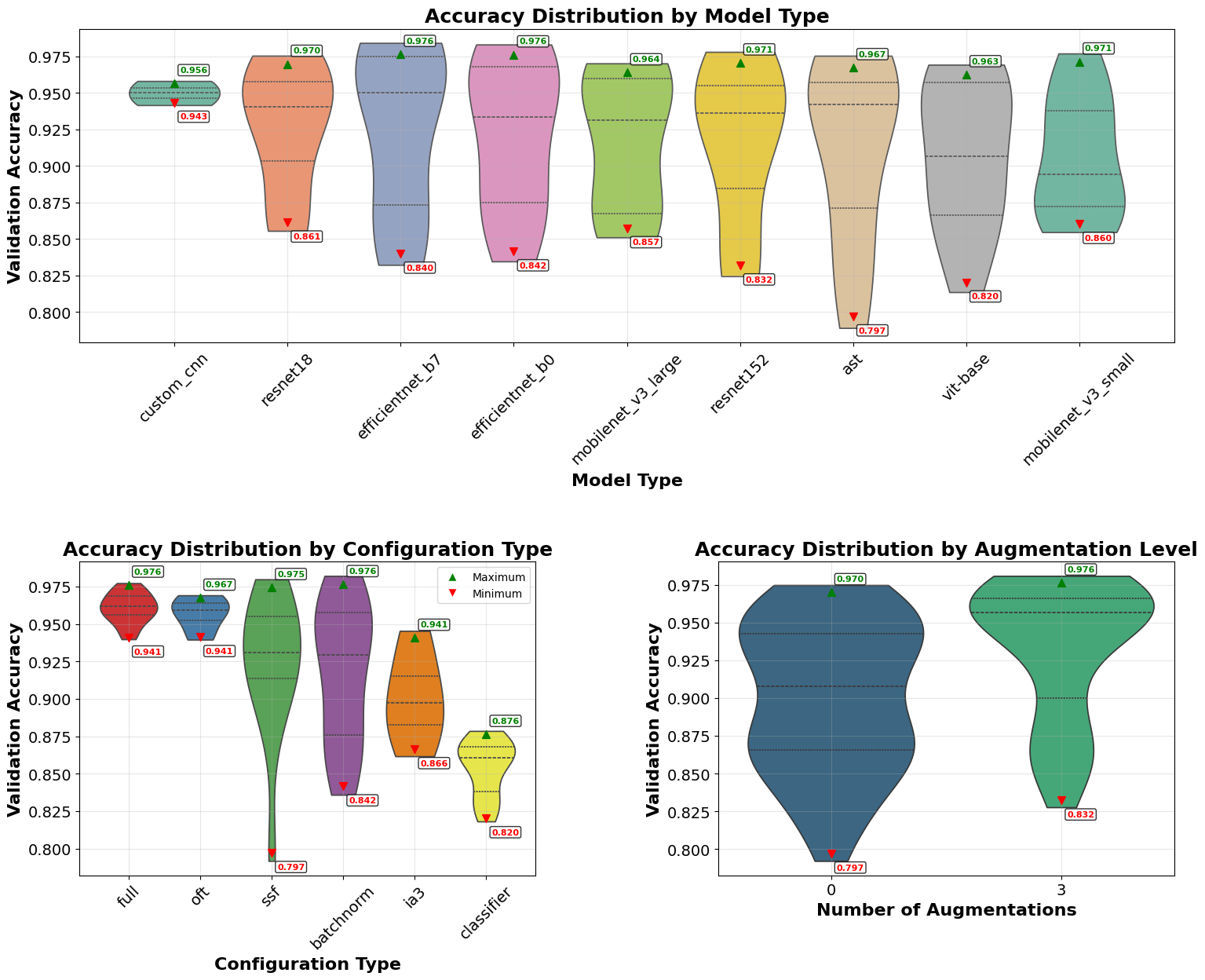}
    \caption{Plots showing Model Validation Accuracies}
    \label{fig:accuracies}
\end{figure}

The experiments tracked loss, accuracies, F1 scores, trainable parameters, and training times. The full training logs are available online \cite{b5}. All nine model architectures (custom CNN, ResNet-18/152, MobileNet-V3 small/large, EfficientNet-B0/B7, ViT-Base, AST) are evaluated for accuracy in Table~\ref{tab:kfold-results} and for training time in Table~\ref{tab:training-table}. Plots showing validation accuracies for representative configurations are presented in Fig.~\ref{fig:accuracies}. There is very little deviation of highest accuracy. Best runs for each model type vary only by 3--4\%. The highest validation accuracy was $97.65\% \pm 0.30$ by EfficientNet-B7 trained using batch-norm fine-tuning and three-fold augmentation. The full results are reported in Table~\ref{tab:kfold-results}.

Importantly, accuracy is not the only metric relevant. Efficiency is crucial for UAV classification, as training models can be computationally expensive and serving models in the field can be limited by battery and onboard compute. We measure the training time, percentage of trainable parameters, and approximate memory footprints to give a clear view of what models most applicable.

\begin{figure}[ht]
    \centering

    \includegraphics[width=1\textwidth]{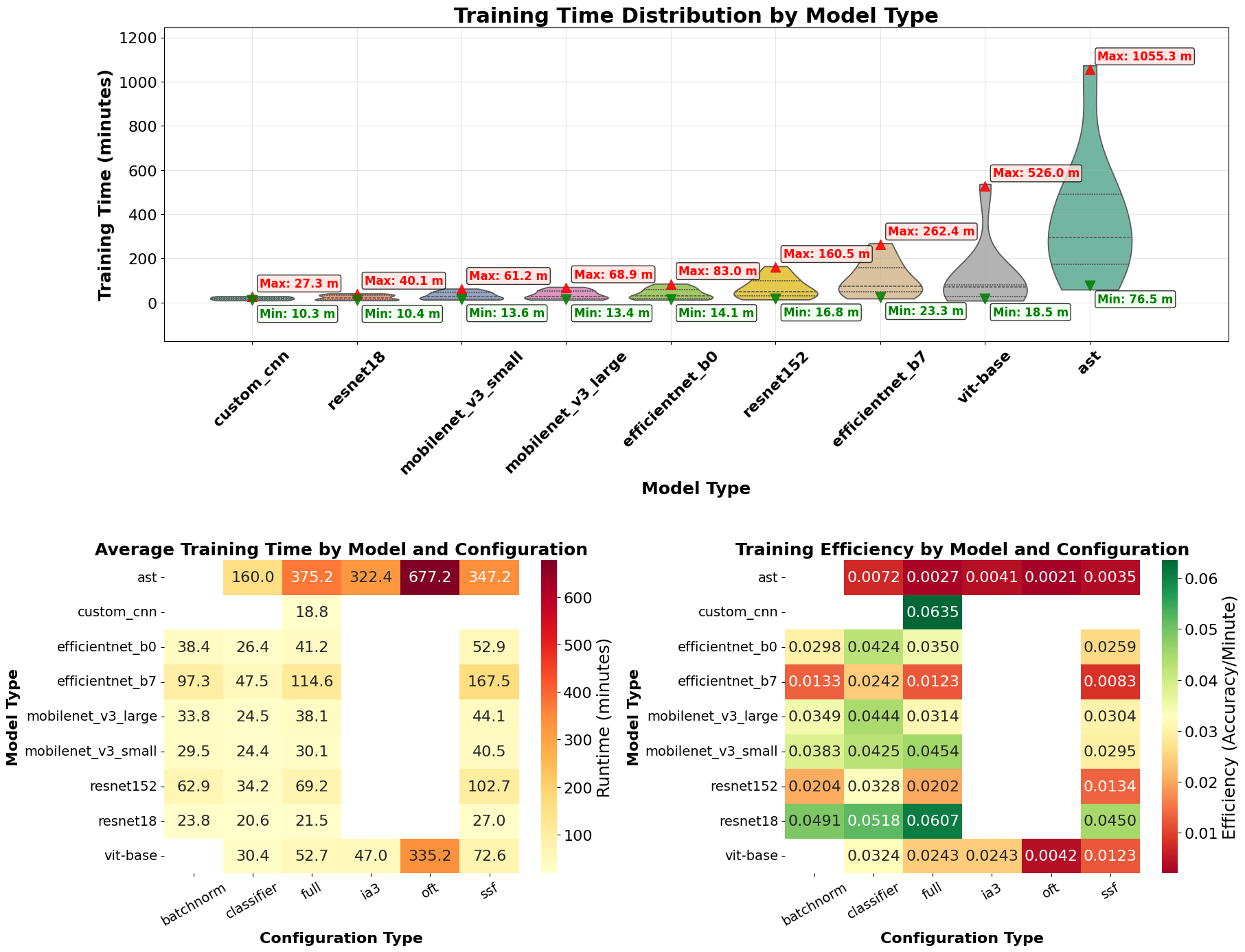}
    \caption{Figures Comparing Model Training Times}
    \label{fig:training-times}
\end{figure}

Figure \ref{fig:training-times} compares the training times between model and PEFT configurations. Note, training times were evaluated using 5-fold cross validation, feature extraction, and augmentations; which can impact the logged times. Efficiency is measured as $\mathrm{Acc/Min} = \mathrm{acc}_{\max} / \bar{t}_{\mathrm{fold}}$, where $\mathrm{acc}_{\max}$ is the best 5-fold validation accuracy for a given (model, PEFT) configuration and $\bar{t}_{\mathrm{fold}}$ is the mean per-fold training wall time in minutes on the hardware described in the Technical Environment subsection. The metric is intended for relative comparison across configurations within this paper rather than as a hardware-independent benchmark. The CNNs scale evenly with time to performance. On average, the AST and ViT models trained for longer, with lower accuracy; making them far less efficient. A comparison of augmentation training times is shown in appendix Fig. \ref{fig:aug-times}. The full table of training times is shown in appendix table \ref{tab:training-table}. The memory footprint is measured in Sec. \ref{sec:footprint}. We now examine which architectures strike the best balance for UAV classification tasks, and why PEFT is of interest.


\begin{sidewaystable}[p]
\centering
\scriptsize
\caption{All 5-Fold Cross Validation Results. Blank entries indicate that a PEFT method is not applicable to that backbone (see Applicability summary in the PEFT subsection).}
\label{tab:kfold-results}
\begin{tabular}{lcccccc}
\hline
\textbf{Model} & \textbf{Full} & \textbf{Classifier} & \textbf{Batch Norm} & \textbf{SSF} & \textbf{Ia3} & \textbf{OFT} \\
\hline
ViT 0 Augs & 95.71\% $\pm$ 0.71\% & 82.03\% $\pm$ 1.06\% & -- & 47.29\% $\pm$ 7.17\% & 86.65\% $\pm$ 1.47\% & 94.13\% $\pm$ 0.86\% \\
ViT 3 Augs & 96.10\% $\pm$ 0.75\% & 86.00\% $\pm$ 1.25\% & -- & 89.32\% $\pm$ 1.60\% & 90.68\% $\pm$ 1.16\% & 96.26\% $\pm$ 0.82\% \\
AST 0 Augs & 96.51\% $\pm$ 1.01\% & 83.36\% $\pm$ 1.22\% & -- & 79.71\% $\pm$ 0.45\% & 88.81\% $\pm$ 1.40\% & 95.61\% $\pm$ 0.72\% \\
AST 3 Augs & 95.74\% $\pm$ 1.01\% & 86.58\% $\pm$ 0.65\% & -- & 94.32\% $\pm$ 1.05\% & 94.07\% $\pm$ 0.79\% & 96.74\% $\pm$ 0.24\% \\
Our CNN 0 Augs & 95.65\% $\pm$ 0.96\% & -- & -- & -- & -- & -- \\
Our CNN 3 Augs & 94.32\% $\pm$ 1.09\% & -- & -- & -- & -- & -- \\
ResNet-18 0 Augs & 95.61\% $\pm$ 1.22\% & 86.13\% $\pm$ 2.34\% & 92.52\% $\pm$ 1.34\% & 91.68\% $\pm$ 1.44\% & -- & -- \\
ResNet-18 3 Augs & 96.97\% $\pm$ 0.68\% & 86.39\% $\pm$ 2.08\% & 95.71\% $\pm$ 1.12\% & 96.00\% $\pm$ 0.43\% & -- & -- \\
ResNet-152 0 Augs & 94.10\% $\pm$ 2.33\% & 83.19\% $\pm$ 2.81\% & 90.23\% $\pm$ 1.72\% & 93.10\% $\pm$ 0.72\% & -- & -- \\
ResNet-152 3 Augs & 95.13\% $\pm$ 1.87\% & 83.23\% $\pm$ 3.01\% & 96.48\% $\pm$ 1.50\% & 97.07\% $\pm$ 0.75\% & -- & -- \\
MobileNet-Small 0 Augs & 95.07\% $\pm$ 0.97\% & 86.87\% $\pm$ 0.55\% & 86.03\% $\pm$ 1.85\% & 87.42\% $\pm$ 2.00\% & -- & -- \\
MobileNet-Small 3 Augs & 97.13\% $\pm$ 0.49\% & 87.32\% $\pm$ 0.60\% & 93.32\% $\pm$ 0.46\% & 91.45\% $\pm$ 3.78\% & -- & -- \\
MobileNet-Large 0 Augs & 96.39\% $\pm$ 0.99\% & 86.81\% $\pm$ 1.68\% & 86.55\% $\pm$ 1.78\% & 91.29\% $\pm$ 1.56\% & -- & -- \\
MobileNet-Large 3 Augs & 96.36\% $\pm$ 1.09\% & 85.74\% $\pm$ 1.76\% & 95.87\% $\pm$ 0.66\% & 94.97\% $\pm$ 0.84\% & -- & -- \\
EfficientNet-B0 0 Augs & \textbf{97.00\% $\pm$ 0.50\%} & 87.65\% $\pm$ 1.54\% & 84.16\% $\pm$ 2.43\% & 91.74\% $\pm$ 0.94\% & -- & -- \\
EfficientNet-B0 3 Augs & 97.61\% $\pm$ 0.44\% & 87.10\% $\pm$ 1.12\% & 94.97\% $\pm$ 1.29\% & 96.71\% $\pm$ 0.72\% & -- & -- \\
EfficientNet-B7 0 Augs & 96.23\% $\pm$ 0.48\% & 85.3\% $\pm$ 1.04\% & 87.97\% $\pm$ 0.60\% & 93.84\% $\pm$ 1.64\% & -- & -- \\
EfficientNet-B7 3 Augs & \textbf{97.58\% $\pm$ 0.29\%} & 84.00\% $\pm$ 0.88\% & \textbf{97.65\% $\pm$ 0.30\%} & \textbf{97.45\% $\pm$ 0.78\%} & -- & -- \\
\hline
\end{tabular}
\end{sidewaystable}




\section{Discussion}
This paper extends prior work on our custom UAV dataset for UAV audio classification. We expanded the methodology of using pre-trained, parameter efficient fine-tuning (PEFT), and data augmentations.

\textbf{RQ1: CNNs vs.\ transformers.} On our 31-class dataset, fine-tuned CNNs outperform fine-tuned transformers in both accuracy and efficiency. The best CNN configuration (EfficientNet-B7 with batch-norm fine-tuning and three-fold augmentation, $97.65\% \pm 0.30$) exceeds the best transformer configuration (AST with OFT and three-fold augmentation, $96.74\% \pm 0.24$) by nearly a percentage point, while training in roughly one-seventh the wall time per fold and using a smaller inference-time memory footprint (243 MB vs.\ 333 MB). Notably, AST is the least efficient model in the sweep despite its close architectural relationship to ViT, likely due to its spectrogram-patch modifications and its longer convergence on small-data tasks. The well-known transformer data-scale advantage is not realized at our 3,100-clip scale; the inductive-bias deficit of transformers on image- and spectrogram-style inputs is not yet compensated by the available data.

\textbf{RQ2: best accuracy-to-efficiency configuration.} Selective batch-norm fine-tuning of EfficientNet-B7 with three-fold augmentation is the strongest combination across our metrics: highest validation accuracy ($97.65\% \pm 0.30$), only 0.49\% of model parameters trained (Table~\ref{tab:trainable-perc}), and a 243 MB inference-time memory footprint (Table~\ref{tab:memory_footprint}). Batch-norm fine-tuning consistently outperformed the additive PEFT methods (SSF, IA3, OFT) on CNN backbones, suggesting that for this task the most useful pretrained signal lives in the convolutional feature extractors and that adapting batch-norm statistics is enough to specialize those features to the UAV distribution. Three-fold augmentation contributed 1--3 percentage points across nearly all model/PEFT pairs (Table~\ref{tab:kfold-results}), making it a strict accuracy win at the cost of roughly a $3\times$ increase in per-epoch wall time.

\textbf{RQ3: within-family backbone scaling.} Scaling the backbone within a family produces different effects across families. For EfficientNet, the smaller B0 reaches its peak ($97.61\% \pm 0.44$) only under full fine-tuning, whereas B7 reaches its peak ($97.65\% \pm 0.30$) under batchnorm fine-tuning, with full fine-tuning a close second. Scaling up the backbone therefore \textit{enables} lightweight PEFT to match full fine-tuning, which is precisely the regime where the title's ``weight'' burden is alleviated. ResNet shows the same pattern at lower absolute accuracy: ResNet-152 with SSF ($97.07\% \pm 0.75$) edges out ResNet-18 with full fine-tuning ($96.97\% \pm 0.68$). MobileNet-V3 is the exception: scaling from small to large under full fine-tuning yields a small decrease in peak accuracy ($97.13\% \pm 0.49$ for small vs.\ $96.36\% \pm 1.09$ for large), suggesting that MobileNet's design is optimized for compute constrained deployment rather than for serving as a base for further PEFT style adaptation. Across families, the consistent finding is that larger backbones make PEFT effective; smaller backbones require full fine-tuning to remain competitive.

Why CNNs outperform transformers in this setting is consistent with three reinforcing factors. First, locality and translation equivariance built into convolutional kernels are well-matched to mel-spectrogram inputs: drone-audio identity lives in time-localized harmonic patterns whose presence matters more than their absolute time index, and CNNs encode exactly that invariance without having to learn it from data. Transformers must instead infer equivalent structure from positional encodings and self-attention, which requires more data than our 3,100-clip dataset supplies. Second, the pretraining each architecture inherits is only weakly aligned with UAV audio. The ViT checkpoint we use is pretrained on ImageNet, and AST extends a ViT backbone with AudioSet pretraining; neither corpus contains drone-specific spectral patterns, so the transfer to UAV audio is shallow and the small-data inductive-bias advantage that CNNs hold is not closed by either pretraining route. Third, the strong showing of batch-norm fine-tuning specifically on the largest CNN we evaluated (EfficientNet-B7) suggests that for a sufficiently large pretrained convolutional feature extractor the per-channel normalization statistics carry most of the burden of adapting ImageNet features to UAV audio: the convolutional features already encode the right invariances, and what needs to change is the distribution-level scale and shift of each channel rather than the features themselves. The lower absolute accuracy of CNN backbones that lack a strong pretrained signal, notably our randomly-initialized custom CNN at $94.32\% \pm 1.09\%$ with three-fold augmentation, is the boundary case that makes the importance of the pretrained features visible.

Several directions extend this work. The most consequential is moving from static evaluation to a continual-learning setting: when a deployed system encounters a new, previously unseen drone, the same PEFT mechanics that make batch-norm fine-tuning of EfficientNet-B7 attractive in our results (fewer than 0.5\% of parameters updated, a small memory footprint, and short retraining time) are exactly what is required to adapt the model on the new platform without curating a fresh full dataset or retraining from scratch. A modestly-sized pretrained model with a PEFT adapter is the natural substrate for that kind of online adaptation. A second direction is multi-modality: integrating visual and radar telemetry into the same classification pipeline so the system can fall back across modalities when one is degraded. A third direction is empirical scaling: as the audio dataset grows beyond 31 classes and 3,100 clips, the boundary at which transformers begin to overtake CNNs should be characterized directly rather than predicted from prior literature, and the dependence of each PEFT method on data scale should be measured. Finally, the strong showing of batch-norm fine-tuning on CNNs invites a theoretical question that this paper does not resolve: how much of the apparent generalization of pretraining-plus-PEFT is feature-level transfer versus distribution-level adaptation, and how does that decomposition shift with model scale?

For UAV audio classification under data scarcity, our results favor lightweight CNN backbones paired with selective fine-tuning over large transformers with full fine-tuning: scaling the method, not the model, gives the best balance of accuracy, training efficiency, and inference-time footprint.

\section{Declarations}
\textbf{Funding:} This work is supported by the South Carolina Research Authority (SCRA)

\textbf{Availability of Data and materials:} The code is available at: \url{https://github.com/AndrewPBerg/UAV_Classification}
\\
The model training logs are available at: \url{https://wandb.ai/andberg9-self/31-class-kfold/table?nw=nwuserandberg9}

\textbf{Conflict of Interest} On behalf of all the authors, the corresponding states there is no conflict of interest.

\textbf{Ethical Approval} Not applicable

\bibliography{Unbearable_Weight}

@inproceedings{b1,
  author    = {A. P. Berg and M. Y. Wang and Q. Zhang},
  title     = {4,500 Seconds: Small data training approaches for deep UAV audio classification},
  booktitle = {Proceedings of the 14th International Conference on Data Science, Technology and Applications (DATA)},
  year      = {2025},
  url       = {https://www.insticc.org/Primoris/Resources/PaperPdf.ashx?idPaper=ScI3f83+ah4=}
}

@article{b2,
  author    = {K. Zaman and M. Sah and C. Direkoglu and M. Unoki},
  title     = {A survey of audio classification using deep learning},
  journal   = {IEEE Access},
  volume    = {11},
  pages     = {106620--106649},
  year      = {2023},
  doi       = {10.1109/ACCESS.2023.3318015}
}

@misc{rethingingCNNs,
      title={Rethinking CNN Models for Audio Classification}, 
      author={Kamalesh Palanisamy and Dipika Singhania and Angela Yao},
      year={2020},
      eprint={2007.11154},
      archivePrefix={arXiv},
      primaryClass={cs.CV},
      url={https://arxiv.org/abs/2007.11154}, 
}

@article{b3,
  author    = {A. F. R. Nogueira and H. S. Oliveira and J. J. M. Machado and J. M. R. S. Tavares},
  title     = {Transformers for urban sound classification---A comprehensive performance evaluation},
  journal   = {Sensors},
  volume    = {22},
  number    = {22},
  pages     = {8874},
  year      = {2022},
  doi       = {10.3390/s22228874}
}

@misc{b4,
  author    = {A. Berg},
  title     = {UAV Classification},
  howpublished = {[Computer software]},
  year      = {2025},
  url       = {https://github.com/AndrewPBerg/UAV_Classification}
}

@misc{b5,
  author    = {A. Berg},
  title     = {Weights \& Biases Project Logs},
  howpublished = {[Online]},
  year      = {2025},
  url       = {https://wandb.ai/andberg9-self/Unbearable-Weight-Results?nw=nwuserandberg9}
}

@misc{b6,
  author    = {A. Paszke and S. Gross and F. Massa and A. Lerer and J. Bradbury and G. Chanan and T. Killeen and Z. Lin and N. Gimelshein and L. Antiga and A. Desmaison and A. K{\"o}pf and E. Yang and Z. DeVito and M. Raison and A. Tejani and S. Chilamkurthy and B. Steiner and L. Fang and J. Bai and S. Chintala},
  title     = {PyTorch: An imperative style, high-performance deep learning library},
  note      = {arXiv preprint arXiv:1912.01703},
  year      = {2019},
  url       = {https://arxiv.org/abs/1912.01703}
}

@misc{b7,
  author    = {{PyTorch Foundation}},
  title     = {PyTorch Documentation},
  year      = {2025},
  howpublished = {[Online]},
  url       = {https://docs.pytorch.org/docs/stable/index.html}
}

@misc{b8,
  author    = {{Hugging Face}},
  title     = {Transformers Documentation},
  year      = {2025},
  howpublished = {[Online]},
  url       = {https://huggingface.co/docs/transformers/en/index}
}

@misc{b9,
  author    = {{Hugging Face}},
  title     = {PEFT Documentation},
  year      = {2025},
  howpublished = {[Online]},
  url       = {https://huggingface.co/docs/peft/en/index}
}

@misc{b10,
  author    = {{Lightning AI}},
  title     = {PyTorch Lightning Documentation},
  year      = {2025},
  howpublished = {[Online]},
  url       = {https://lightning.ai/docs/pytorch/stable/}
}

@misc{b11,
  author    = {{Weights \& Biases}},
  title     = {Weights \& Biases Documentation},
  year      = {2025},
  howpublished = {[Online]},
  url       = {https://docs.wandb.ai/}
}

@misc{b12,
  author    = {{Docker}},
  title     = {Docker Documentation},
  year      = {2025},
  howpublished = {[Online]},
  url       = {https://docs.docker.com/}
}

@misc{b13,
  author    = {{PyTorch Foundation}},
  title     = {TorchAudio Documentation},
  year      = {2025},
  howpublished = {[Online]},
  url       = {https://docs.pytorch.org/audio/stable/index.html}
}

@misc{b14,
  author    = {{Audiomentations}},
  title     = {Audiomentations Documentation},
  year      = {2025},
  howpublished = {[Online]},
  url       = {https://iver56.github.io/audiomentations/}
}

@misc{b15,
  author    = {{TorchMetrics}},
  title     = {TorchMetrics Documentation},
  year      = {2025},
  howpublished = {[Online]},
  url       = {https://lightning.ai/docs/torchmetrics/stable/}
}

@misc{b16,
  author    = {{Matplotlib Developers}},
  title     = {Matplotlib Documentation},
  year      = {2025},
  howpublished = {[Online]},
  url       = {https://matplotlib.org/stable/index.html}
}

@misc{b17,
  author    = {{scikit-learn Developers}},
  title     = {scikit-learn Documentation},
  year      = {2025},
  howpublished = {[Online]},
  url       = {https://scikit-learn.org/stable/}
}

@misc{b18,
  author    = {{NumPy Developers}},
  title     = {NumPy Documentation},
  year      = {2025},
  howpublished = {[Online]},
  url       = {https://numpy.org/doc/}
}

@misc{b19,
  author    = {{Librosa Developers}},
  title     = {Librosa Documentation},
  year      = {2025},
  howpublished = {[Online]},
  url       = {https://librosa.org/doc/latest/index.html}
}

@misc{b21,
  author    = {F. Wolf-Monheim},
  title     = {Spectral and rhythm features for audio classification with deep convolutional neural networks},
  note      = {arXiv preprint arXiv:2410.06927},
  year      = {2024},
  month     = {Oct},
  url       = {https://arxiv.org/abs/2410.06927}
}

@inproceedings{b22,
  author    = {Y. Wang and Z. Chu and I. Ku and E. C. Smith and E. T. Matson},
  title     = {A large-scale UAV audio dataset and audio-based UAV classification using CNN},
  booktitle = {Proceedings of the Sixth IEEE International Conference on Robotic Computing (IRC)},
  address   = {Naples, Italy},
  pages     = {186--189},
  year      = {2022},
  month     = {Dec},
  doi       = {10.1109/IRC55401.2022.00039}
}

@article{b23,
  author    = {S. Kümmritz},
  title     = {The sound of surveillance: Enhancing machine learning-driven drone detection with advanced acoustic augmentation},
  journal   = {Drones},
  volume    = {8},
  number    = {3},
  pages     = {105},
  year      = {2024},
  month     = {Mar},
  doi       = {10.3390/drones8030105},
  url       = {https://doi.org/10.3390/drones8030105}
}

@misc{b24,
  author    = {A. P. Berg},
  title     = {Augmentations Colab Notebook},
  year      = {2025},
  howpublished = {[Online]},
  url       = {https://colab.research.google.com/drive/1bl4RTQd7ENnMYEc4thwBwtocF-q1NYp2?usp=sharing}
}

@article{b25,
  author    = {M. Iman and H. R. Arabnia and K. Rasheed},
  title     = {A review of deep transfer learning and recent advancements},
  journal   = {Technologies},
  volume    = {11},
  number    = {2},
  pages     = {40},
  year      = {2023},
  month     = {Feb},
  doi       = {10.3390/technologies11020040}
}

@misc{b29,
  author    = {G. Alain and Y. Bengio},
  title     = {Understanding intermediate layers using linear classifier probes},
  note      = {arXiv preprint arXiv:1610.01644},
  year      = {2018},
  month     = {Nov},
  url       = {https://arxiv.org/abs/1610.01644}
}

@misc{b30,
  author    = {Z. Qiu and W. Liu and H. Feng and Y. Xue and Y. Feng and Z. Liu and D. Zhang and A. Weller and B. Schölkopf},
  title     = {Controlling text-to-image diffusion by orthogonal finetuning},
  note      = {arXiv preprint arXiv:2306.07280},
  year      = {2023},
  month     = {Jun},
  url       = {https://arxiv.org/abs/2306.07280}
}

@misc{b31,
  author    = {H. Liu and D. Tam and M. Muqeeth and J. Mohta and T. Huang and M. Bansal and C. Raffel},
  title     = {Few-shot parameter-efficient fine-tuning is better and cheaper than in-context learning},
  note      = {arXiv preprint arXiv:2205.05638},
  year      = {2022},
  month     = {May},
  url       = {https://arxiv.org/abs/2205.05638}
}

@misc{b32,
  author    = {D. Lian and D. Zhou and J. Feng and X. Wang},
  title     = {Scaling \& shifting your features: A new baseline for efficient model tuning},
  note      = {arXiv preprint arXiv:2210.08823},
  year      = {2022},
  month     = {Oct},
  url       = {https://arxiv.org/abs/2210.08823}
}

@misc{b33,
  author    = {A. P. Berg},
  title     = {UAV Classification SSF PEFT code},
  year      = {2025},
  howpublished = {[Online]},
  url       = {https://github.com/AndrewPBerg/UAV_Classification/blob/master/src/models/ssf_adapter.py}
}

@misc{b34,
  author    = {J. Frankle and D. J. Schwab and A. S. Morcos},
  title     = {Training BatchNorm and only BatchNorm: On the expressive power of random features in CNNs},
  note      = {arXiv preprint arXiv:2003.00152},
  year      = {2020},
  month     = {Mar},
  url       = {https://arxiv.org/abs/2003.00152}
}

@misc{b35,
  author    = {A. Dosovitskiy and others},
  title     = {An image is worth 16x16 words: Transformers for image recognition at scale},
  note      = {arXiv preprint arXiv:2010.11929},
  year      = {2020},
  month     = {Oct},
  url       = {https://arxiv.org/abs/2010.11929}
}

@misc{b36,
  author    = {K. Palanisamy and D. Singhania and A. Yao},
  title     = {Rethinking CNN models for audio classification},
  note      = {arXiv preprint arXiv:2007.11154},
  year      = {2020},
  month     = {Jul},
  url       = {https://arxiv.org/abs/2007.11154}
}

@misc{b37,
  author    = {Y. Gong and Y.-A. Chung and J. Glass},
  title     = {AST: Audio Spectrogram Transformer},
  note      = {arXiv preprint arXiv:2104.01778},
  year      = {2021},
  month     = {Apr},
  url       = {https://arxiv.org/abs/2104.01778}
}

@article{Wang_dissertation,
author = "Yaqin Wang",
title = "{A LARGE-SCALE UAV AUDIO DATASET AND AUDIO-BASED UAV CLASSIFICATION USING CNN}",
year = "2023",
month = "7",
url = "https://hammer.purdue.edu/articles/thesis/_strong_A_LARGE-SCALE_UAV_AUDIO_DATASET_AND_AUDIO-BASED_UAV_CLASSIFICATION_USING_CNN_strong_/23696391",
doi = "10.25394/PGS.23696391.v1"
}

@misc{b38,
  author    = {O. Russakovsky and others},
  title     = {ImageNet large scale visual recognition challenge},
  note      = {arXiv preprint arXiv:1409.0575},
  year      = {2014},
  month     = {Sep},
  url       = {https://arxiv.org/abs/1409.0575}
}

@inproceedings{b39,
  author    = {J. F. Gemmeke and others},
  title     = {Audio Set: An ontology and human-labeled dataset for audio events},
  booktitle = {Proceedings of the IEEE International Conference on Acoustics, Speech and Signal Processing (ICASSP)},
  address   = {New Orleans, LA, USA},
  pages     = {776--780},
  year      = {2017},
  doi       = {10.1109/ICASSP.2017.7952261},
  url       = {https://ieeexplore.ieee.org/document/7952261}
}

@misc{b40,
  author    = {K. He and X. Zhang and S. Ren and J. Sun},
  title     = {Deep residual learning for image recognition},
  note      = {arXiv preprint arXiv:1512.03385},
  year      = {2015},
  month     = {Dec},
  url       = {https://arxiv.org/abs/1512.03385}
}

@misc{b41,
  author    = {{TorchVision}},
  title     = {ResNet --- TorchVision main documentation},
  howpublished = {[Online]},
  url       = {https://pytorch.org/vision/main/models/resnet.html}
}

@misc{b42,
  author    = {A. G. Howard and others},
  title     = {MobileNets: Efficient convolutional neural networks for mobile vision applications},
  note      = {arXiv preprint arXiv:1704.04861},
  year      = {2017},
  month     = {Apr},
  url       = {https://arxiv.org/abs/1704.04861}
}

@misc{b43,
  author    = {A. Howard and others},
  title     = {Searching for MobileNetV3},
  note      = {arXiv preprint arXiv:1905.02244},
  year      = {2019},
  month     = {May},
  url       = {https://arxiv.org/abs/1905.02244}
}

@misc{b44,
  author    = {{TorchVision}},
  title     = {MobileNet V3 --- TorchVision main documentation},
  howpublished = {[Online]},
  url       = {https://pytorch.org/vision/main/models/mobilenetv3.html}
}

@misc{b45,
  author    = {M. Tan and Q. V. Le},
  title     = {EfficientNet: Rethinking model scaling for convolutional neural networks},
  note      = {arXiv preprint arXiv:1905.11946},
  year      = {2019},
  month     = {May},
  url       = {https://arxiv.org/abs/1905.11946}
}

@misc{b46,
  author    = {{TorchVision}},
  title     = {EfficientNet --- TorchVision main documentation},
  howpublished = {[Online]},
  url       = {https://pytorch.org/vision/main/models/efficientnet.html}
}

@misc{b47,
  author    = {V. Semenyuk and I. Kurmashev and A. Lupidi and D. Alyoshin and L. Kurmasheva and A. Cantelli-Forti},
  title     = {Advance and Refinement: The Evolution of UAV Detection and Classification Technologies},
  note      = {arXiv preprint arXiv:2409.05985},
  year      = {2024},
  month     = {Sep},
  url       = {https://arxiv.org/abs/2409.05985}
}

@misc{b48,
  author    = {Md. M. Rahman and S. Siddique and M. Kamal and R. H. Rifat and K. D. Gupta},
  title     = {UAV (Unmanned Aerial Vehicles): Diverse Applications of UAV Datasets in Segmentation, Classification, Detection, and Tracking},
  note      = {arXiv preprint arXiv:2409.03245},
  year      = {2024},
  month     = {Sep},
  url       = {https://arxiv.org/abs/2409.03245}
}

@inproceedings{b50,
  author    = {M. Y. Wang and D. C. Ramirez and E. Noonan and M. Linn and Q. Zhang},
  title     = {A Comprehensive Dataset and Visualization Tool for Drone Acoustic Signatures},
  booktitle = {Proceedings of the 2024 Artificial Intelligence x Humanities, Education, and Art (AIxHEART)},
  address   = {Laguna Hills, CA, USA},
  pages     = {13--17},
  year      = {2024},
  month     = {Sep},
  doi       = {10.1109/AIxHeart62327.2024.00009},
  url       ={https://www.researchgate.net/publication/388017272\_A\_Comprehensive\_Dataset\_and\_Visualization\_Tool\_for\_Drone\_Acoustic\_Signatures}
}

@article{b51,
  author    = {M. Y. Wang and Z. Chu and I. Ku and E. C. Smith and E. T. Matson},
  title     = {A 15-Category Audio Dataset for Drones and an Audio-Based UAV Classification Using Machine Learning},
  journal   = {International Journal of Semantic Computing},
  volume    = {18},
  number    = {2},
  pages     = {257--272},
  year      = {2024},
  doi       = {10.1142/S1793351X24300048},
  url       = {https://doi.org/10.1142/S1793351X24300048}
}

@article{mu2021environmental,
  title        = {Environmental sound classification using temporal-frequency attention based convolutional neural network},
  author       = {Mu, W. and Yin, B. and Huang, X. and Xu, J. and Du, Z.},
  journal      = {Scientific Reports},
  volume       = {11},
  pages        = {21552},
  year         = {2021},
  doi          = {10.1038/s41598-021-01045-4},
}

@misc{vaswani2023,
      title={Attention Is All You Need}, 
      author={Ashish Vaswani and Noam Shazeer and Niki Parmar and Jakob Uszkoreit and Llion Jones and Aidan N. Gomez and Lukasz Kaiser and Illia Polosukhin},
      year={2023},
      eprint={1706.03762},
      archivePrefix={arXiv},
      primaryClass={cs.CL},
      url={https://arxiv.org/abs/1706.03762}, 
}

@misc{sun2017,
      title={Revisiting Unreasonable Effectiveness of Data in Deep Learning Era}, 
      author={Chen Sun and Abhinav Shrivastava and Saurabh Singh and Abhinav Gupta},
      year={2017},
      eprint={1707.02968},
      archivePrefix={arXiv},
      primaryClass={cs.CV},
      url={https://arxiv.org/abs/1707.02968}, 
}


\begin{appendices}

\section{Data Collection}\label{secA1}
Full Tabular Description \ref{tab:uav_data_31} for the custom UAV audio dataset including collection site, manufacturer, and model name.
\begin{sidewaystable}[p]
\centering
\scriptsize
\caption{UAV Audio Dataset: 31 Classes with Collection Sites}
\label{tab:uav_data_31}
\begin{tabular}{lllrrl}
\hline
\textbf{Manufacture} & \textbf{Model} & \textbf{Drone Type} & \textbf{Number of Files} & \textbf{Duration (sec)} & \textbf{Collection Site} \\
\hline
Self-build & David Tricopter & Outdoor & 100 & 500 & Columbus, IN \\
Self-build & PhenoBee & Outdoor & 100 & 500 & West Lafayette, IN \\
Autel & Evo 2 Pro & Outdoor & 100 & 500 & New Richmond, IN \\
DJI & Avata & Outdoor & 100 & 500 & Charleston, SC \\
DJI & FPV & Outdoor & 100 & 500 & Charleston, SC \\
DJI & Matrice 200 & Outdoor & 100 & 500 & West Lafayette, IN \\
DJI & Matrice 200 V2 & Outdoor & 100 & 500 & New Richmond, IN \\
DJI & Matrice 600p & Outdoor & 100 & 500 & New Richmond, IN \\
DJI & Mavic Air 2 & Outdoor & 100 & 500 & New Richmond, IN \\
DJI & Mavic Mini 1 & Outdoor & 100 & 500 & New Richmond, IN \\
DJI & Mini 2 & Outdoor & 100 & 500 & New Richmond, IN \\
DJI & Mini 3 & Outdoor & 100 & 500 & Charleston, SC \\
DJI & Mini 3 Pro & Outdoor & 100 & 500 & Charleston, SC \\
DJI & Mavic 2 Pro & Outdoor & 100 & 500 & New Richmond, IN \\
DJI & Mavic 2s & Outdoor & 100 & 500 & New Richmond, IN \\
DJI & Phantom 2 & Outdoor & 100 & 500 & New Richmond, IN \\
DJI & Phantom 4 & Outdoor & 100 & 500 & New Richmond, IN \\
DJI & Tello & Indoor & 100 & 500 & Charleston, SC \\
DJI & RoboMaster TT Tello & Indoor & 100 & 500 & New Richmond, IN \\
Hasakee & Q11 & Indoor & 100 & 500 & West Lafayette, IN \\
Holystone & HS210 & Indoor & 100 & 500 & Charleston, SC \\
Hover & X1 & Outdoor & 100 & 500 & Charleston, SC \\
Syma & X5SW & Indoor & 100 & 500 & West Lafayette, IN \\
Syma & X5UW & Indoor & 100 & 500 & West Lafayette, IN \\
Syma & X8SW & Indoor & 100 & 500 & West Lafayette, IN \\
Syma & X20 & Indoor & 100 & 500 & West Lafayette, IN \\
Syma & X20P & Indoor & 100 & 500 & West Lafayette, IN \\
Syma & X26 & Indoor & 100 & 500 & West Lafayette, IN \\
Swellpro & Splash 3 Plus & Outdoor & 100 & 500 & New Richmond, IN \\
Yuneec & Typhoon H Plus & Outdoor & 100 & 500 & New Richmond, IN \\
UDI RC & U46 & Outdoor & 100 & 500 & West Lafayette, IN \\
\hline
\multicolumn{2}{r}{\textbf{Total}} & & \textbf{3,100} & \textbf{15,500} & \\
\hline
\end{tabular}
\end{sidewaystable}

\section{Memory Footprint}\label{sec:footprint}
The memory footprint is measured as the inference-time footprint of any model specific. This is a key metric for edge-devices and overall scale.
\[
\text{Memory}_{\text{MB}} = \frac{\text{TotalParams} \times b}{1024 \times 1024}
\]

\quad \text{where } b \text{ is the number of bytes per parameter.}
so for float32 precision,
\[
b = 4 \;\;\Rightarrow\;\;
\text{Memory}_{\text{MB}} = \frac{\text{TotalParams} \times 4}{1024 \times 1024}
\]

\begin{figure}[ht]
    \centering

    \includegraphics[width=1\textwidth]{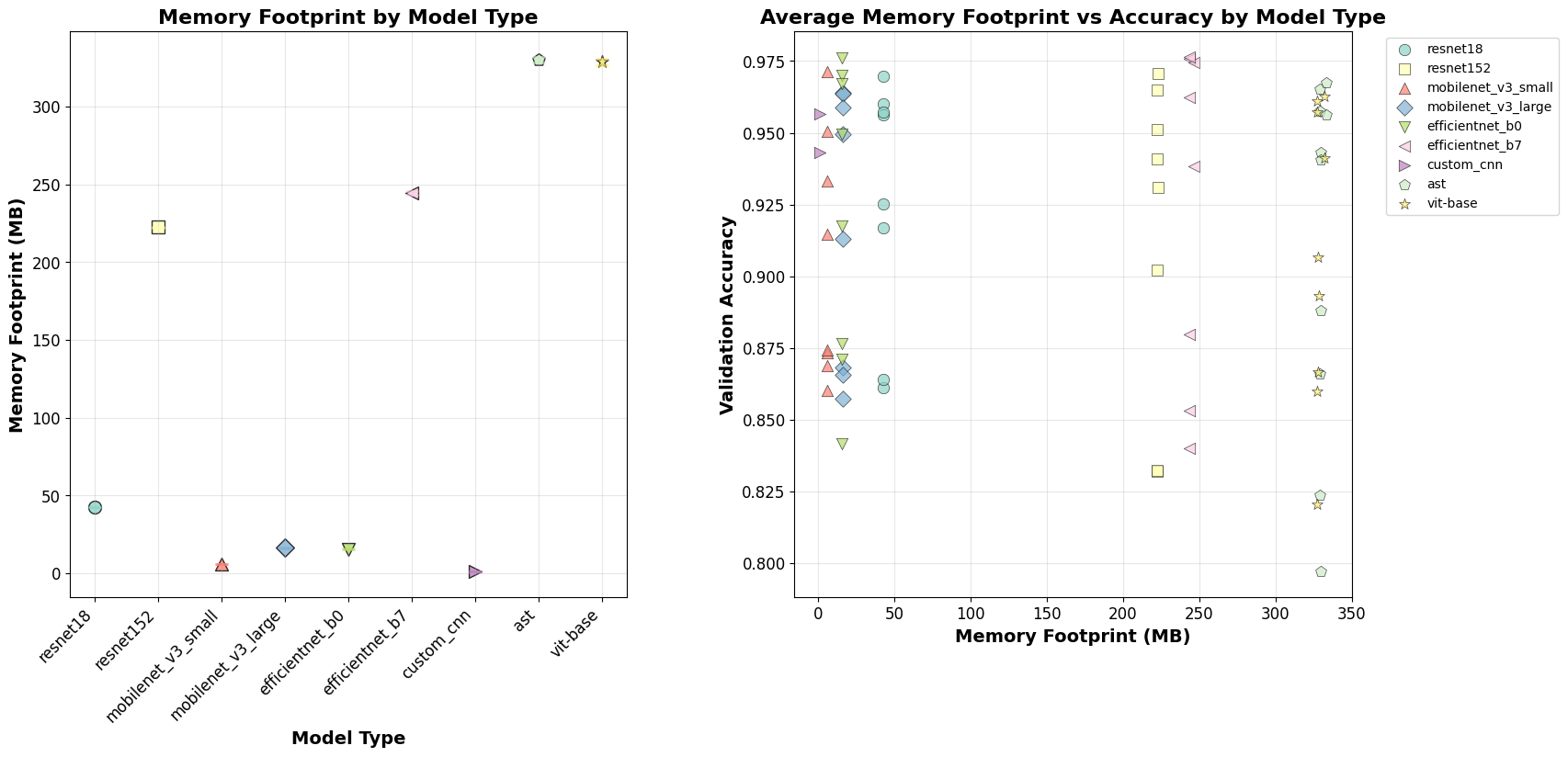}
    \caption{Plots showing Model Training Times}
    \label{fig:footprint}
\end{figure}

As shown in Fig. \ref{fig:footprint} there is little correlation between model accuracy and memory footprint.
Table \ref{tab:memory_footprint} shows the full breakdown between model and PEFT configuration. Note that additive PEFT methods like SSF, IA3, and OFT slightly raise the memory footprint.

\begin{table}[h]
\centering
\begin{tabular}{lcccccc}
\hline
\textbf{Model} & \textbf{Full} & \textbf{Classifier} & \textbf{BatchNorm} & \textbf{SSF} & \textbf{Ia3} & \textbf{OFT} \\
\hline
ResNet18            & 42.67 & 42.67 & 42.67 & 42.74 & -     & - \\
ResNet152           & 222.02 & 222.02 & 222.02 & 223.17 & -     & - \\
MobileNetV3-Small   & 5.91 & 5.91 & 5.91 & 6.03 & -     & - \\
MobileNetV3-Large   & 16.18 & 16.18 & 16.18 & 16.41 & -     & - \\
EfficientNet-B0     & 15.44 & 15.44 & 15.44 & 15.83 & -     & - \\
EfficientNet-B7     & 243.63 & 243.63 & 243.63 & 246.56 & -     & - \\
Custom CNN          & 1.12 & -     & -     & -     & -     & - \\
AST                 & 328.88 & 328.88 & -     & 329.66 & 329.29 & 333.40 \\
ViT-Base            & 327.39 & 327.39 & -     & 328.17 & 327.79 & 331.91 \\
\hline
\end{tabular}
\caption{Memory footprint (MB) of different models under various PEFT configurations.}
\label{tab:memory_footprint}
\end{table}




\section{Training Time vs. Augmentations}\label{appdx-augs}
\begin{figure}[ht]
    \centering

    \includegraphics[width=1\textwidth]{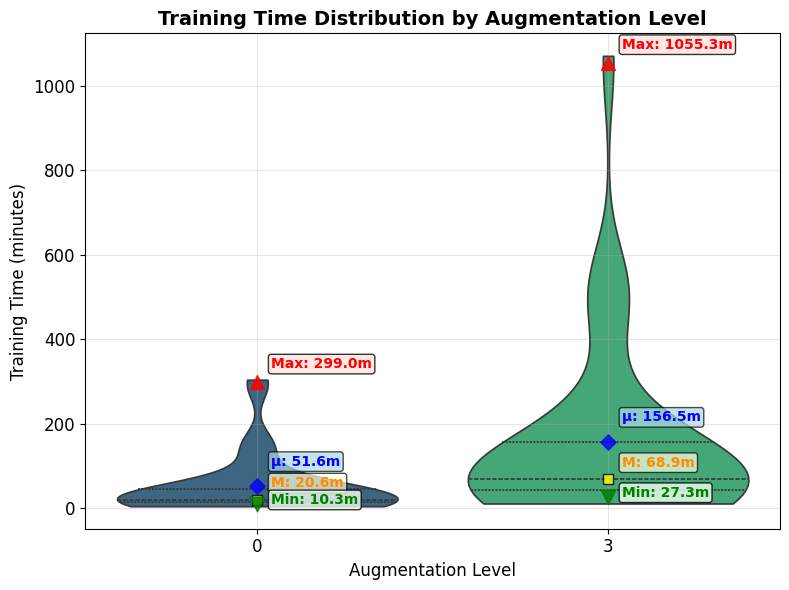}
    \caption{Training Times vs. Data Augmentations}
    \label{fig:aug-times}
\end{figure}

Data augmentations significantly increase training times as shown in Fig. \ref{fig:aug-times}.
\section{Training Time Table}\label{secE1}
Table \ref{tab:training-table} shows the average training time of different models. Blank entries indicate PEFT incompatibility.

\begin{table}[h]
\centering
\begin{tabular}{lcccccc}
\hline
\textbf{Model} & \textbf{Full} & \textbf{Classifier} & \textbf{BatchNorm} & \textbf{SSF} & \textbf{Ia3} & \textbf{OFT} \\
\hline
ResNet18            & 21.54 & 20.58 & 23.84 & 27.00 & -     & - \\
ResNet152           & 69.24 & 34.22 & 62.85 & 102.68 & -     & - \\
MobileNetV3-Small   & 30.13 & 24.36 & 29.50 & 40.52 & -     & - \\
MobileNetV3-Large   & 38.12 & 24.53 & 33.76 & 44.14 & -     & - \\
EfficientNet-B0     & 41.17 & 26.43 & 38.37 & 52.95 & -     & - \\
EfficientNet-B7     & 114.64 & 47.49 & 97.32 & 167.55 & -     & - \\
Custom CNN          & 18.77 & -     & -     & -     & -     & - \\
AST                 & 375.18 & 160.01 & -     & 347.18 & 322.36 & 677.18 \\
ViT-Base            & 52.66 & 30.35 & -     & 48.88 & 46.00 & 335.24 \\
\hline
\end{tabular}
\caption{Average training runtime (minutes) of different models under various PEFT configurations.}
\label{tab:training-table}
\end{table}

\section{Trainable Parameter Percentage Table}\label{secE2}
Table \ref{tab:trainable-perc} shows the percentage of trainable parameters for each model type. Full fine-tuning is training all parameters, thus is 100\%.

\begin{sidewaystable}[p]
\centering
\scriptsize
\caption{Trainable Parameter Percentage for Each Model and PEFT Configuration}\label{tab:trainable-perc}
\begin{tabular}{lcccccc}
\hline
\textbf{Model} & \textbf{Full (\%)} & \textbf{Classifier (\%)} & \textbf{Batch Norm (\%)} & \textbf{SSF (\%)} & \textbf{Ia3 (\%)} & \textbf{OFT (\%)} \\
\hline

ViT-Base & 100.00 & 0.03 & -- & 0.24 & 0.12 & 1.36 \\
AST & 100.00 & 0.03 & -- & 0.24 & 0.13 & 1.36 \\
Custom CNN & 100.00 & -- & -- & -- & -- & -- \\
ResNet-18 & 100.00 & 0.14 & 0.09 & 0.17 & -- & -- \\
ResNet-152 & 100.00 & 0.11 & 0.26 & 0.52 & -- & -- \\
MobileNet-V3 Small & 100.00 & 40.19 & 0.78 & 2.03 & -- & -- \\
MobileNet-V3 Large & 100.00 & 29.94 & 0.58 & 1.43 & -- & -- \\
EfficientNet-B0 & 100.00 & 0.98 & 1.04 & 2.48 & -- & -- \\
EfficientNet-B7 & 100.00 & 0.12 & 0.49 & 1.19 & -- & -- \\

\hline
\end{tabular}
\end{sidewaystable}




\end{appendices}

\end{document}